\documentclass[12pt,draftclsnofoot, perreview, onecolumn]{IEEEtran}%Prof. Ma's template
\usepackage{balance}
\usepackage{color}
\usepackage{url}
\usepackage{threeparttable}
\usepackage{amsfonts}
\usepackage{amsmath}
\usepackage{mathrsfs}
\usepackage{multirow}
\usepackage{amssymb}
\usepackage{mdwmath}
\usepackage{ifpdf}
\usepackage[dvips]{graphicx}%GRAPHICS RELATED PACKAGES
\graphicspath{{../}}
\DeclareGraphicsExtensions{.pdf,.jpeg,.png,.eps}

\ifCLASSOPTIONcompsoc
  \usepackage[tight,normalsize,sf,SF]{subfigure}%SUBFIGURE PACKAGES
\else
  \usepackage[tight,footnotesize]{subfigure}%SUBFIGURE PACKAGES
\fi

\newcommand{\mathbfit}[1]{\mbox{\boldmath$#1$\unboldmath}}
\newtheorem{algorithm}{\textbf{Algorithm}}
\newtheorem{example}{\textbf{Example}}

\newcommand{\Ref}[1]{(\ref{#1})}

\newcommand{\noisy}{noisy}

\newcommand{\Prob}[1]{\Pr\left\{#1\right\}}

\newcommand{\vecleft}{\left(}
\newcommand{\vecright}{\right)}

\newcommand{\phaseOnePe}{{p_{\rm genie}}}
\newcommand{\bscPe}{p_{\rm flip}}

\newcommand{\sumLayerBit}{r}

\newcommand{\errBitNum}{r}

\newcommand{\CarN}{N}
\newcommand{\CarK}{K}
\newcommand{\CarBlocks}{B}
\newcommand{\Qfun}[1]{{\rm Q}\left(#1\right)}
\newcommand{\kIdx}{g}
\newcommand{\nIdx}{h}
\newcommand{\basicN}{n}
\newcommand{\basicK}{k}

\newcommand{\nWeightIdx}{h}
\newcommand{\kWeightIdx}{g}
\newcommand{\target}{\rm target}

\ifCLASSOPTIONonecolumn
\newcommand{\figwidth}{0.65\textwidth}

\newcommand{\vpscaefigure}{\vspace{0.0cm}}

\newcommand{\vpscaesection}{\vspace{0.0cm}}

\fi

\ifCLASSOPTIONtwocolumn
\newcommand{\figwidth}{0.48\textwidth}

\newcommand{\vpscaefigure}{\vspace{0.0cm}}

\newcommand{\vpscaesection}{\vspace{0.0cm}}

\fi
\begin{document}

%% Paper Title
\title{Spatial Coupling of Generator Matrix: A General Approach to Design of Good Codes at a Target BER}

%% Author
\author{Chulong~Liang,
        Xiao~Ma,~\IEEEmembership{Member,~IEEE,}
        Qiutao~Zhuang,
        and~Baoming~Bai,~\IEEEmembership{Member,~IEEE}% <-this % stops a space
%\thanks{Manuscript received $\cdots$}% <-this % stops a space
\thanks{This work was supported by the 973 Program (No.2012CB316100) and the NSF~(No.61172082) of China.}% <-this % stops a space
%This paper was presented in part at the IEEE International Symposium on Information Theory, 2013.
\thanks{Chulong~Liang, Xiao~Ma, Qiutao~Zhuang are with the Department of Electronics and Communication
Engineering, Sun Yat-sen University, Guangzhou 510006, China (e-mail: lchul@mail2.sysu.edu.cn, zhuangqt@mail2.sysu.edu.cn, maxiao@mail.sysu.edu.cn).}
\thanks{Baoming~Bai is with State Lab. of ISN, Xidian University, Xi'an 710071, Shaanxi, China (e-mail: bmbai@mail.xidian.edu.cn).}}

%% Create the title:
\maketitle

%% Abstract:
\begin{abstract}
%Block Markov superposition transmission~(BMST) is a simple coding scheme for constructing big convolutional codes by introducing memory between the basic codewords. As a result, BMST can be viewed as the spatial coupling of the generator matrix of the basic code.
For any given short code~(referred to as the {\em basic code}), block Markov superposition transmission~(BMST) provides a simple way to obtain {\em predictable} extra coding gain by spatial coupling the generator matrix of the basic code.
This paper presents a systematic design methodology for BMST systems to approach the channel capacity at any given target bit-error-rate~(BER) of interest. To simplify the design, we choose the basic code as the Cartesian product of a short block code. The encoding memory is then inferred from the genie-aided lower bound according to the performance gap of the short block code to the corresponding Shannon limit at the target BER. In addition to the sliding-window decoding algorithm, we propose to perform one more phase decoding to {remove residual~(rare) errors}. A new technique that assumes a \noisy~genie is proposed to upper bound the performance. Under some mild assumptions, these genie-aided bounds can be used to predict the performance of the proposed two-phase decoding algorithm in the extremely low BER region. Using the Cartesian product of a repetition code as the basic code, we construct a BMST system with an encoding memory 30 whose performance at the BER of $10^{-15}$ can be predicted within one dB away from the Shannon limit over the binary-input additive white Gaussian noise channel (BI-AWGNC).
\end{abstract}

\begin{IEEEkeywords}
Big convolutional codes, block Markov superposition transmission~(BMST), capacity-approaching codes, genie-aided bounds, spatial coupling, two-phase decoding.
\end{IEEEkeywords}

\section{Introduction}\label{sec:Introduction}
Turbo codes~\cite{Berrou93}, which were invented by Berrou~\emph{et al} in 1993, can achieve with an interleaver of size $65536$ a bit-error-rate (BER) of $10^{-5}$ at the signal-to-noise ratio (SNR) about $0.5$~dB away from the Shannon limit. Since a BER of $10^{-5}$ meets the requirement of most applications, turbo codes have been adopted by many wireless communication standards~\cite{Costello07}, such as UMTS, CDMA2000, IEEE 802.16, etc. However, different systems may have different performance requirements. For example, an optical recording system requires a BER of $10^{-12}$ or lower~\cite{Marchant90}, while a magnetic storage system requires an even lower BER~\cite{Sripimanwat05}, typically, of $10^{-15}$. When the BER drops from $10^{-5}$ down to $10^{-10}$, the original turbo code has more than $5$ dB coding gain loss due to the existence of the error floor~\cite{Costello09}. It is also the error floor phenomenon that has restricted the applications of turbo codes to the scenarios where a very low BER is required.

Low-density parity-check~(LDPC) codes, which were invented by Gallager~\cite{Gallager63} and rediscovered after the invention of turbo codes, may be promising for applications where a very low BER is required. {In~\cite{Richardson01}, Richardson~\emph{et al} proposed density evolution~(DE) to design irregular LDPC codes with thresholds very close to the corresponding capacities. However, in order to approach the threshold, the block length should be large enough.} To the best of our knowledge, no randomly constructed codes were reported in the literature to perform within one dB away from the Shannon limit at the BER of around $10^{-15}$. More promising than randomly constructed codes, algebraically constructed codes~\cite{Zhang10,Huang12} have been shown by simulation with field-programmable gate array (FPGA) decoders to have no visible error floor down to the BER of $10^{-12}$ or even $10^{-14}$, where a code of rate $0.8752$ performs $1.6$ dB away from the Shannon limit at the BER of $10^{-14}$.

Recently, a coding scheme called block Markov superposition transmission~(BMST) of short codes~(referred to as \emph{basic codes}) was proposed~\cite{Ma13,Ma13x}, which has a good performance over the binary-input additive white Gaussian noise channel~(BI-AWGNC). A BMST system is indeed a convolutional~(linear or nonlinear) code~\cite{Ma13x}, as is similar to a convolutional LDPC code~\cite{Felstrom99,Tanner04,Lentmaier10,Pusane11}.
However, its parity-check matrix plays no role in either the construction or the decoding. As a sub-codeword of BMST is the superposition of several consecutive {\em interleaved} basic codewords, the BMST can be viewed as the spatial coupling~\cite{Kudekar11} of the generator matrix of the basic code.
The most important feature of the BMST system is the derived genie-aided performance lower bound, which is simple but relates the coding gain to the encoding memory.
%The genie-aided lower bound is also useful to justify the near-optimality of the proposed sliding-window decoding~(SWD) algorithm in the low error rate region~\cite{Ma13}.
The genie-aided lower bound is also useful to justify the near-optimality of the proposed sliding-window decoding~(SWD) algorithm in the low error rate region~\cite{Ma13x}. This motivates us to construct good codes with a predictable error floor at a target BER, especially at some extremely low BER~(say $10^{-15}$).
%\color{red}This provides us a simple way to construct good codes with a predictable error floor at a target BER.

{In this paper, we propose a general procedure to construct BMST systems to approach the channel capacity at a target BER of interest. To analyze the performance in the extremely low error rate region, we propose a new bounding technique that assumes a genie-aided decoder. Along with the genie-aided upper bound is a two-phase decoding (TPD) algorithm, where the phase-I decoding is the SWD algorithm serving as a genie and the phase-II is the well-known minimum Euclidean distance decoding algorithm removing residual rare errors.
%Simulation results around the BER of $10^{-5}$ show that the performance of the TPD algorithm matches well with the genie-aided upper bound.
A BMST system is constructed by taking as the basic code the Cartesian product of the simplest repetition code. With an encoding memory $m = 30$ and a decoding delay $d = 60$, the performance upper bound of the TPD algorithm has a BER lower than $10^{-15}$ at $E_b/N_0 = 0.5$~dB, which is within one dB away from the Shannon limit.

The rest of this paper is organized as follows. In Section~\ref{sec:ReviewBMST}, we review the BMST and present a general procedure to design BMST systems to approach the corresponding Shannon limits. In Section~\ref{sec:TwoPhaseDecoding}, we derive a new bounding technique for the BMST based on a noisy genie-aided decoder and present the TPD algorithm for the BMST, whose performance can be upper bounded by the genie-aided bound in the low error rate region. Simulation results for a BMST system constructed at a target BER of $10^{-15}$ is also presented in Section~\ref{sec:TwoPhaseDecoding}. Section~\ref{sec:Conclusion} concludes this paper.}
%The rest of this paper is organized as follows. In Section~\ref{sec:ReviewBMST}, we review the BMST and present a general procedure to design BMST systems to approach the corresponding Shannon limits. In Section~\ref{sec:TwoPhaseDecoding}, we present the TPD algorithm for the BMST and derive a genie-aided bound for the phase-II decoding in the low error rate region. In Section~\ref{sec:NumericalResults}, we give construction examples together with simulation results. Section~\ref{sec:Conclusion} concludes this paper.
%In Section~\ref{sec:UpperBound}, we derive a new bounding technique for the BMST based on a \noisy~genie-aided decoder.

\section{Design of BMST with a Target BER}\label{sec:ReviewBMST}
\subsection{Encoding of BMST}
\begin{figure}[t]
   \centering
   \includegraphics[width=\figwidth]{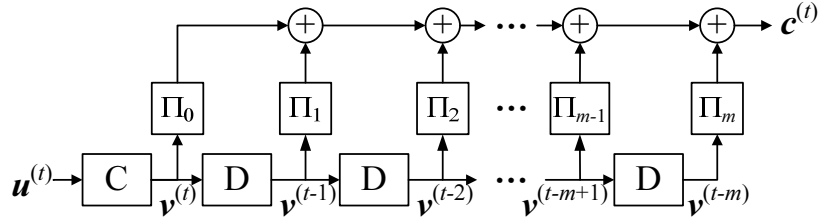}
   %\caption{System Model of the BMST system over an AWGN channel using BPSK signalling.}
   \caption{The encoding diagram of a BMST system with memory $m$.}
   \label{fig:SystemModel}\vpscaefigure
\end{figure}
Consider a BMST system of an encoding memory $m$ using a binary basic code $\mathscr{C}[\basicN,\basicK]$ of length $\basicN$ and dimension $\basicK$, see Fig.~\ref{fig:SystemModel} for reference. Let $\mathbfit{u} = \left(\mathbfit{u}^{(0)}, \cdots, \mathbfit{u}^{(L-1)}\right)$ be $L$ blocks of data to be transmitted, where $\mathbfit{u}^{(t)}=\vecleft{u}^{(t)}_{0}, {u}^{(t)}_{1}, \cdots, {u}^{(t)}_{\basicK-1}\vecright \in \mathbb{F}_2^\basicK (0 \leq t \leq L-1)$. To terminate the encoding process,  $m$ extra blocks of all-zero vectors $\mathbfit{u}^{(t)}\!=\!\mathbf{0}\!\in\!\mathbb{F}_2^\basicK(L\!\leq\!t\!\leq\!L\!+\!m\!-\!1)$ are padded. For $t=0,1,\cdots,L+m-1$, $\mathbfit{u}^{(t)}$ is first encoded into an intermediate codeword $\mathbfit{v}^{(t)}=\vecleft{v}^{(t)}_{0}, \cdots, {v}^{(t)}_{\basicN-1}\vecright \in \mathbb{F}_{2}^{\basicN}$ by the basic encoder, which is then used to compute the sub-codeword $\mathbfit{c}^{(t)}\!=\!\vecleft{c}^{(t)}_{0},\!\cdots,\!{c}^{(t)}_{\basicN\!-\!1}\!\vecright\!\in\!\mathbb{F}_{2}^{\basicN}$ for transmission as
\begin{eqnarray}\label{eq:Superposition}
  \mathbfit{c}^{(t)} &=& \mathbfit{v}^{(t)}\mathbfit{\varPi}_{0} + \mathbfit{v}^{(t-1)}\mathbfit{\varPi}_{1} + \cdots + \mathbfit{v}^{(t-m)}\mathbfit{\varPi}_{m},
\end{eqnarray}
where $\mathbfit{v}^{(t)}$ for $t<0$ are initialized as the all-zero vectors and $\mathbfit{\varPi}_{i}\left(0\leq i \leq m\right)$ are $m+1$ permutation matrices of size $\basicN \times \basicN$. For the BI-AWGNC, $\mathbfit{\varPi}_{0}$ can be set as the identity matrix. It can be seen that, in the case when the basic code is itself a conventional convolutional code and $m=1$, the BMST system is reduced to a hyperimposed convolutional code~\cite{Cheng96}.

Denote $\mathbfit{w}^{(t, i)} = \mathbfit{v}^{(t)} \mathbfit{\varPi}_i$ for $0 \leq i \leq m$. From the encoding process, we can see that the sub-block $\mathbfit{u}^{(t)}$ is encoded into $\left(\mathbfit{w}^{(t, 0)}, \mathbfit{w}^{(t, 1)}, \cdots, \mathbfit{w}^{(t, m)}\right)$ and ``mixed" into $m+1$ coded sub-blocks $\left(\mathbfit{c}^{(t)}, \mathbfit{c}^{(t+1)}, \cdots, \mathbfit{c}^{(t+m)}\right)$. Ignoring the effects of boundary~(initialization and termination), a typical coded sub-block $\mathbfit{c}^{(t)}$ is a superposition of $m+1$ sub-blocks, that is, $\mathbfit{c}^{(t)} = \sum_{i=0}^{m} \mathbfit{w}^{(t-i, i)}$.
The superposition introduces memory among codewords of the basic code, resembling the spatial coupling LDPC codes~\cite{Kudekar11}. However, it is the generator matrix instead of the parity-check matrix that is coupled~\cite{Ma13x}.
%The superposition process introduces memory between $m+1$ adjacent codewords of the basic code, resulting in the spatial coupling of the generator matrix of the basic code~\cite{Kudekar11}, which can also be seen from the generator matrix of the BMST~\cite{Ma13x}.

\subsection{Decoding of BMST}
We assume that $\mathbfit{c}^{(t)}\left(0\leq t \leq L+m-1\right)$ are transmitted over the BI-AWGNC. Firstly, $\mathbfit{c}^{(t)}$ is mapped into a bipolar signal sequence $\mathbfit{x}^{(t)} = \left( {x}^{(t)}_{0}, \cdots, {x}^{(t)}_{n-1} \right)$ for $0\leq t \leq L+m-1$, where ${x}^{(t)}_{j} = \left(-1\right)^{{c}^{(t)}_{j}}$ for $0\!\leq\!j\!\leq\!\basicN\!-\!1$. For convenience, the mapping is written in a compact form $\mathbfit{x}^{(t)} = \left(-1\right)^{\mathbfit{c}^{(t)}}$. At the receiver, the noisy received vector $\mathbfit{y}^{(t)}=\vecleft{y}^{(t)}_{0}, \cdots, {y}^{(t)}_{\basicN-1}\vecright$ can be expressed as
\begin{equation}\label{eq:reveiving}
y^{(t)}_{j} = {x}^{(t)}_{j} + z_{j}^{(t)}
\end{equation}
for $0\!\leq\!j\!\leq\!\basicN\!-\!1$ and $0\!\leq\!t\!\leq\!L\!+\!m\!-\!1$, where $z_{j}^{(t)}\left(0\!\leq\!j\!\leq\!\basicN\!-\!1,\!0\!\leq\!t\!\leq\!L\!+\!m\!-\!1\right)$ are independent and identically distributed~(i.i.d.) with Gaussian distribution $\mathcal{N}\left( 0, \sigma^2 \right)$. Considering only the constraint on the modulation and the channel, the \emph{a posteriori} probabilities~(APPs) input to the decoder can be computed as
\begin{equation}\label{eq:bitAPP}
\Prob{\!C^{(t\!)}_j\!=\!u | {y}^{(t\!)}_{j}\!}\!=\!
\alpha_j \exp\!\left\{\!-\frac{\!\left(y_j^{(\!t\!)}\!-\!\left(-1\right)^{u}\right)^2}{2 \sigma^2}\!\right\},\!u\!\in\!\mathbb{F}_2,
\end{equation}
for $0\!\leq\!j\!\leq\!\basicN\!-\!1$ and $0 \leq t \leq L+m-1$, where $C^{(t)}_{j}$ is the random variable corresponding to ${c}^{(t)}_{j}$ and $\alpha_j$ is the normalized factor to ensure that $\sum_{u \in \mathbb{F}_2}\!\Prob{\!C^{(t\!)}_j\!=\!u | {y}^{(t\!)}_{j}}\!=\!1$.

\begin{figure}[t]
   \centering
   \includegraphics[height=\figwidth,angle=270]{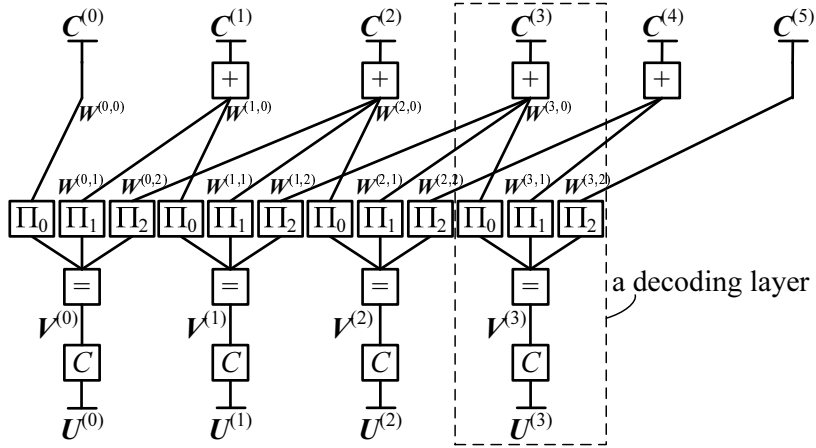}
   \caption{The normal graph for a BMST system with $L=4$ and $m=2$.}
   \label{fig:decoder}\vpscaefigure
\end{figure}
At time $t$, the sliding-window decoding~(SWD) algorithm, which is an iterative message passing/processing algorithm over the corresponding normal graph of the BMST system~\cite{Ma13}, is performed with a decoding delay $d$ and a maximum iteration number $I_{\max}$ to recover $\mathbfit{u}^{(t-d)}$ by taking $\Prob{\!C^{(t-i\!)}_j\!=\!u | {y}^{(t-i\!)}_{j}\!}, u \in \mathbb{F}_2\left(0\leq i \leq d, 0\leq j \leq \basicN-1\right)$ as inputs.  Fig.~\ref{fig:decoder} shows the normal graph of a BMST system with $L\!=\!4$ and $m\!=\!2$.

\subsection{A General Procedure to Design BMST}
It can be seen that the main components of the BMST system include a basic code and at most $m+1$ interleavers. It has been pointed out in~\cite{Ma13} that any code with fast encoding and efficient soft-in-soft-out~(SISO) decoding algorithm can be applied to BMST systems. For interleavers, we have found by simulations that randomly generated interleavers perform already well. The issue is, given a basic code, how to choose the encoding memory. Fortunately, we have the following simple performance lower bound.

Let $p_b = f_{\rm basic}(\gamma_b)$ be the BER performance function of the basic code $\mathscr{C}\left[\basicN, \basicK\right]$, where $p_b$ is the BER and $\gamma_b \stackrel{\Delta}{=} E_b/N_0$ in dB. Let $p_b = f_{\rm BMST}(\gamma_b)$ be the BER performance function for the corresponding BMST system with memory $m$. By assuming a genie-aided decoder, we have~\cite{Ma13x}
\begin{equation}\label{lowerbound}
    f_{\rm BMST}(\gamma_b) \geq f_{\rm basic}\left(\gamma_b\!+\!10\log_{10}\left(m\!+\!1\right)\right).
    %\!-\!10\log_{10}(1\!+\!m/L)).
\end{equation}
In other words, the maximum extra coding gain over the basic code can be $10\log_{10}(m+1)$ dB in the low error rate region for large $L$.

{Now assume that we want to construct a good code with a given rate to approach the Shannon limit at a target BER~(denoted by $p_{\rm target}$).} Then we can design a BMST system following the general procedure as described below.

%\begin{process}
{\bf \em A General Procedure of Designing BMST}\label{pcs:DesignBMST}
\begin{enumerate}
  \item Take a code with the given rate as the basic code. Typically, we can take either a convolutional code with a short constraint length or a Cartesian product of a short block code. In order to approach the channel capacity, we set the code length $n \geq 10000$ in our simulations;
  \item Find the performance curve $f_{\rm basic}\left(\gamma_b\right)$ of the basic code. From this curve, find the required $E_b/N_0$ to achieve the target BER. That is, find $\gamma_{\target}$ such that $f_{\rm basic}(\gamma_{\target}) \leq p_{\target}$;
  \item Find the Shannon limit for the code rate, denoted by $\gamma_{\lim}$;
  \item Determine the encoding memory %The encoding memory is then determined
        by $10\log_{10}(m+1) \geq \gamma_{\target} - \gamma_{\lim}$. That is,
        \begin{equation}\label{eq:ComputeMemory}
        m = \left\lceil 10^{\frac{\gamma_{\target} - \gamma_{\lim}}{10}}-1 \right\rceil,
        \end{equation}
        where $\left\lceil x \right\rceil$ stands for the minimum integer greater than or equal to $x$;
  \item Generate $m+1$ interleavers randomly.
\end{enumerate}
%\end{process}

The above procedure requires no optimization and hence can be easily implemented. The only issue is how to get the performance curve $f_{\rm basic}\left(\gamma_b\right)$ as required by the determination of $\gamma_{\target}$. For the target BER of around $10^{-5}$, we may use simulation, while for the target BER of around $10^{-15}$, we need an analytic~(computable) form either for $f_{\rm basic}(\gamma_b)$ itself or for some upper bound on $f_{\rm basic}(\gamma_b)$. This becomes simple for the Cartesian product of short block codes.

\subsection{Cartesian Product of Short Block Codes}
Let $\mathscr{C}\left[\CarN, \CarK\right]$ be a short binary block code. We can use its $B$-fold Cartesian product, denoted by $\mathscr{C}\left[\CarN, \CarK\right]^\CarBlocks$, as the basic code  $\mathscr{C}\left[\basicN, \basicK\right]$, where $\basicN\!=\!\CarN\CarBlocks$ and $\basicK\!=\!\CarK\CarBlocks$. The use of the Cartesian product of a short block code allows us to implement in a parallel manner both the encoding algorithm and the decoding algorithm, which may be attractive for hardware implementation. More importantly, the BER performance function of the code $\mathscr{C}\left[\CarN, \CarK\right]^{\CarBlocks}$ is the same as the short code $\mathscr{C}\left[\CarN, \CarK\right]$, which simplifies the code design. Let the input-output weight enumerating function~(IOWEF) of the short code $\mathscr{C}\left[\CarN, \CarK\right]$ be given as
\begin{equation}\label{eq:IOWEF_C}
A \left( X, Y \right) \triangleq \sum_{\kIdx=0}^{\CarK}\sum_{\nIdx=0}^{\CarN} A_{\kIdx,\nIdx} X^{\kIdx} Y^{\nIdx},
\end{equation}
where $X$, $Y$ are two dummy variables and $A_{\kIdx,\nIdx}$ denotes the number of codewords having a Hamming weight $\nIdx$ when the corresponding input information sequence having a Hamming weight $\kIdx$. The BER performance $f_{\rm basic}\left(\gamma_b\right)$ of the basic code $\mathscr{C}\left[\basicN, \basicK\right]$ over the BI-AWGNC can be bounded by the union bound as~\cite{Sason06}
\begin{equation}\label{eq:UnionBound_C}
p_b = f_{\rm basic}\left(\gamma_b\right) \leq \sum_{\kIdx=1}^{\CarK}\sum_{\nIdx=1}^{\CarN}\frac{\kIdx}{\CarK}A_{\kIdx,\nIdx} \Qfun{\sqrt{2\nIdx\cdot\frac{\CarK}{\CarN}\cdot 10^{\scriptstyle\frac{\gamma_b}{10}}}},
\end{equation}
where
\begin{equation}\label{eq:Qfunction}
\Qfun{x} = \frac{1}{\sqrt{2\pi}}\int_{x}^{\infty}\exp\left\{-\frac{t^2}{2}\right\}dt.
\end{equation}
Then $\gamma_{\target}$ can be determined by
\begin{eqnarray}\label{eq:ChooseMemory}
p_{\small \rm target} = \sum\limits_{\kIdx=1}^{\CarK}\sum\limits_{\nIdx=1}^{\CarN}\frac{\kIdx}{\CarK}A_{\kIdx,\nIdx} \Qfun{\sqrt{2\nIdx\cdot\frac{\CarK}{\CarN}\cdot 10^{\scriptstyle\frac{\gamma_{\target}}{10}}}}
\end{eqnarray}
using bisection search.
\begin{example}[Determination of encoding memory]%[Constructing BMST with repetition codes]
\label{ex:BMSTwithRC}
We construct BMST systems with different code rates to approach the corresponding Shannon limits at several target BERs over the BI-AWGNC. {We choose five code rates $1/8, 1/4 , 1/2, 3/4,$ and $7/8$. For rates $1/8, 1/4,$ and $1/2$, the Cartesian product of repetition codes~(RCs) are used as basic codes, while for rates $3/4$ and $7/8$, the Cartesian product of single parity-check~(SPC) codes are used as basic codes.
As an example, the detailed construction procedure for rate $1/2$ is given in the following.
%Here, we give a detailed construction procedure for the case with rate $1/2$, where the Cartesian product of RC $[2,1]^B$ is used as the basic code.
}
\begin{enumerate}
  \item Take the Cartesion product of the RC $[2,1]^B$ as the basic code. Set $B=5000$, i.e. the basic code length $n=10000$.
  \item The IOWEF of the RC $[2,1]$ is
        \begin{equation}\label{eq:IOWEF_RC}
            A\left(X,Y\right) = 1 + XY^2.
        \end{equation}
        The performance function of the RC $[2,1]^B$ is the same as the union bound, which is given by
        \begin{equation}\label{eq:performance_RC}
           p_b = f_{\rm basic}\left(\gamma_b\right) = \Qfun{\sqrt{2 \cdot 10^{\scriptstyle\frac{\gamma_b}{10}}}}.
        \end{equation}
        So the required $E_b/N_0$ to achieve the target BERs $10^{-3}, 10^{-5}, 10^{-6},$ and $10^{-15}$ are $6.79$~dB, $9.59$~dB, $10.53$~dB, and $14.99$~dB, respectively.
        %From this curve, find the required $E_b/N_0$ to achieve the target BER. That is, find $\gamma_{\target}$ such that $f_{\rm basic}(\gamma_{\target}) \leq p_{\target}$;
  \item The Shannon limit for rate $1/2$ is $\gamma_{\lim} = 0.19$~dB;
  \item The encoding memories for the target BERs $10^{-3}, 10^{-5}, 10^{-6},$ and $10^{-15}$ computed from~(\ref{eq:ComputeMemory}) are $4,8,10,$ and $30$, respectively;
  \item Generate $m+1$ interleavers of length $n = 10000$ uniformly at random, where $m=5,9,11,$ and $31$ as required by the target BERs $10^{-3}, 10^{-5}, 10^{-6},$ and $10^{-15}$, respectively. Once these interleavers are generated, they are fixed in our simulations.
        %Generate $5,9,11,$ and $31$ interleavers of length $10000$ uniformly at random for the target BERs $10^{-3}, 10^{-5}, 10^{-6},$ and $10^{-15}$, respectively.
\end{enumerate}
%The IOWEF of the RC $[2,1]$ is
%\begin{equation}\label{eq:IOWEF_RC}
%    A\left(X,Y\right) = 1 + XY^2.
%\end{equation}
%The performance function of the RC $[2,1]^B$ is the same as the union bound, which is given by
%\begin{equation}\label{eq:performance_RC}
%p_b = f_{\rm basic}\left(\gamma_b\right) = \Qfun{\sqrt{2 \cdot 10^{\scriptstyle\frac{\gamma_b}{10}}}}.
%\end{equation}
%For the target BER $p_{\rm target}=10^{-3}$, the Shannon limit of rate $1/2$ is $\gamma_{\lim} = 0.11$, so the memory required is $m=4$. Similarly, for $p_{\rm target} = 10^{-5}$ and $10^{-6}$, the memories required are $m=8$ and $10$, respectively. In addition, we construct a BMST system with the RC~$[2,1]^B$ as the basic code at an extremely low target BER $p_{\rm target} = 10^{-15}$, where the memory required is $m=30$.

The encoding memories required for all chosen settings are shown in Table~\ref{tab:MemoryRequired4Example}. As expected, the lower the target BER is, the larger the memory is required.
%\begin{table}[t]
%\caption{The encoding memories required to approach the Shannon limit for different target BERs using the BMST of RC $[2,1]^B$\label{tab:MemoryRequired}}
%\centering
%\begin{tabular}{l|ccc}
%  \hline
%  \hline
%  % after \\: \hline or \cline{col1-col2} \cline{col3-col4} ...
%  Target BER $p_{\target}$ & $10^{-3}$ & $10^{-5}$ & $10^{-15}$ \\
%  \hline
%  Target SNR $\gamma_{\target}$ (dB) & 6.79 & 9.59 & 14.99 \\
%  Shannon limit $\gamma_{\lim}$ (dB) & 0.11 & 0.19 & 0.19 \\
%  Gap $\delta_{\gamma}=\gamma_{\target}-\gamma_{\lim}$ (dB) & 6.68 & 9.40 & 14.80 \\
%  Memory required $m$ & 4 & 8 & 30 \\
%  \hline
%  \hline
%\end{tabular}
%\end{table}
{\color{red}
\begin{table}[t]
\caption{The encoding memories required to approach the corresponding Shannon limits using BMST systems for different code rates at given target BERs\label{tab:MemoryRequired4Example}}
\centering
\ifCLASSOPTIONonecolumn
\begin{tabular}{cccccccc}
\fi
\ifCLASSOPTIONtwocolumn
\begin{tabular}{p{1.6cm}p{0.6cm}p{0.6cm}p{1.0cm}p{1.6cm}p{0.2cm}}
\fi
  \hline
  \hline
  \scriptsize Basic codes & \scriptsize $p_{\target}$ & \scriptsize $\gamma_{\target}$ (dB) & \scriptsize $\gamma_{\lim}$ (dB) & \scriptsize $\gamma_{\target}\!-\!\gamma_{\lim}$ (dB) & \scriptsize $m$ \\
  \hline
  \scriptsize RC $[8,1]^{1250}$ & \scriptsize $10^{-3}$ & \scriptsize $6.79$ & \scriptsize $-1.21$ & \scriptsize $8.00$ & \scriptsize $6$ \\
  \scriptsize RC $[8,1]^{1250}$ & \scriptsize $10^{-6}$ & \scriptsize $10.53$ & \scriptsize $-1.21$ & \scriptsize $11.74$ & \scriptsize $14$ \\
  \scriptsize RC $[4,1]^{2500}$ & \scriptsize $10^{-3}$ & \scriptsize $6.79$ & \scriptsize $-0.79$ & \scriptsize $7.58$ & \scriptsize $5$ \\
  \scriptsize RC $[4,1]^{2500}$ & \scriptsize $10^{-6}$ & \scriptsize $10.53$ & \scriptsize $-0.79$ & \scriptsize $11.32$ & \scriptsize $13$ \\
  \scriptsize RC $[2,1]^{5000}$ & \scriptsize $10^{-3}$ & \scriptsize $6.79$ & \scriptsize $0.19$ & \scriptsize $6.60$ & \scriptsize $4$ \\
  \scriptsize RC $[2,1]^{5000}$ & \scriptsize $10^{-5}$ & \scriptsize $9.59$ & \scriptsize $0.19$ & \scriptsize $9.40$ & \scriptsize $8$ \\
  \scriptsize RC $[2,1]^{5000}$ & \scriptsize $10^{-6}$ & \scriptsize $10.53$ & \scriptsize $0.19$ & \scriptsize $10.34$ & \scriptsize $10$ \\
  \scriptsize RC $[2,1]^{5000}$ & \scriptsize $10^{-15}$ & \scriptsize $14.99$ & \scriptsize $0.19$ & \scriptsize $14.80$ & \scriptsize $30$ \\
  \scriptsize SPC $[4,3]^{2500}$ & \scriptsize $10^{-3}$ & \scriptsize $5.86$ & \scriptsize $1.63$ & \scriptsize $4.23$ & \scriptsize $2$ \\
  \scriptsize SPC $[4,3]^{2500}$ & \scriptsize $10^{-6}$ & \scriptsize $9.15$ & \scriptsize $1.63$ & \scriptsize $7.52$ & \scriptsize $5$ \\
  \scriptsize SPC $[8,7]^{1250}$ & \scriptsize $10^{-3}$ & \scriptsize $5.75$ & \scriptsize $2.84$ & \scriptsize $2.91$ & \scriptsize $1$ \\
  \scriptsize SPC $[8,7]^{1250}$ & \scriptsize $10^{-6}$ & \scriptsize $8.77$ & \scriptsize $2.84$ & \scriptsize $5.93$ & \scriptsize $3$ \\
  %\scriptsize SPC $[10,9]$ & \scriptsize $10^{-15}$ & \scriptsize $12.72$ & \scriptsize $3.20$ & \scriptsize $9.52$ & \scriptsize $8$ \\
  %\scriptsize RM $[16,5]$ & \scriptsize $10^{-15}$ & \scriptsize $11.37$ & \scriptsize $-0.57$ & \scriptsize $ 11.94$ & \scriptsize $15$ \\
  %\scriptsize Extended Hamming $[8,4]$ & \scriptsize $10^{-15}$ & \scriptsize $12.25$ & \scriptsize $0.19$ & \scriptsize $ 12.06$ & \scriptsize $16$ \\
  \hline
  \hline
\end{tabular}
\end{table}
}
\end{example}

\subsection{Numerical Results}
\begin{figure}[t]
   \centering
   \includegraphics[width=\figwidth]{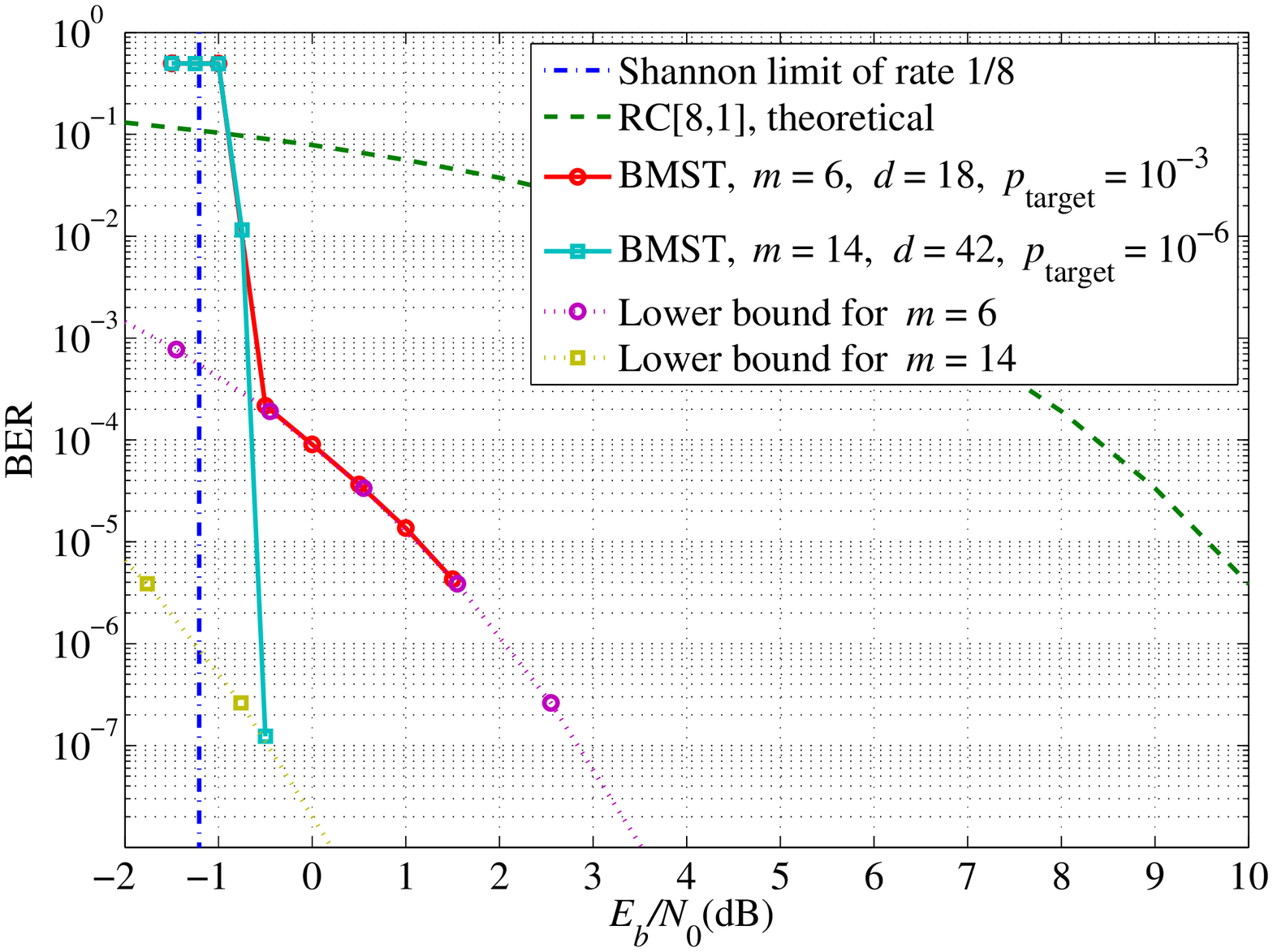}
    \caption{Performance of the BMST systems with the RC $[8,1]^{1250}$ as the basic code. The target BERs are $10^{-3}$ and $10^{-6}$. The systems encode $L=100000$ sub-blocks of data and decode with the SWD algorithm of a maximum iteration $I_{\max}=18$.}
    %and a stopping threshold $\delta_h=10^{-5}$, where encoding memories and the decoding delays are specified in the legend.}
   \label{fig:rc81m3}
\end{figure}

\begin{figure}[t]
   \centering
   \includegraphics[width=\figwidth]{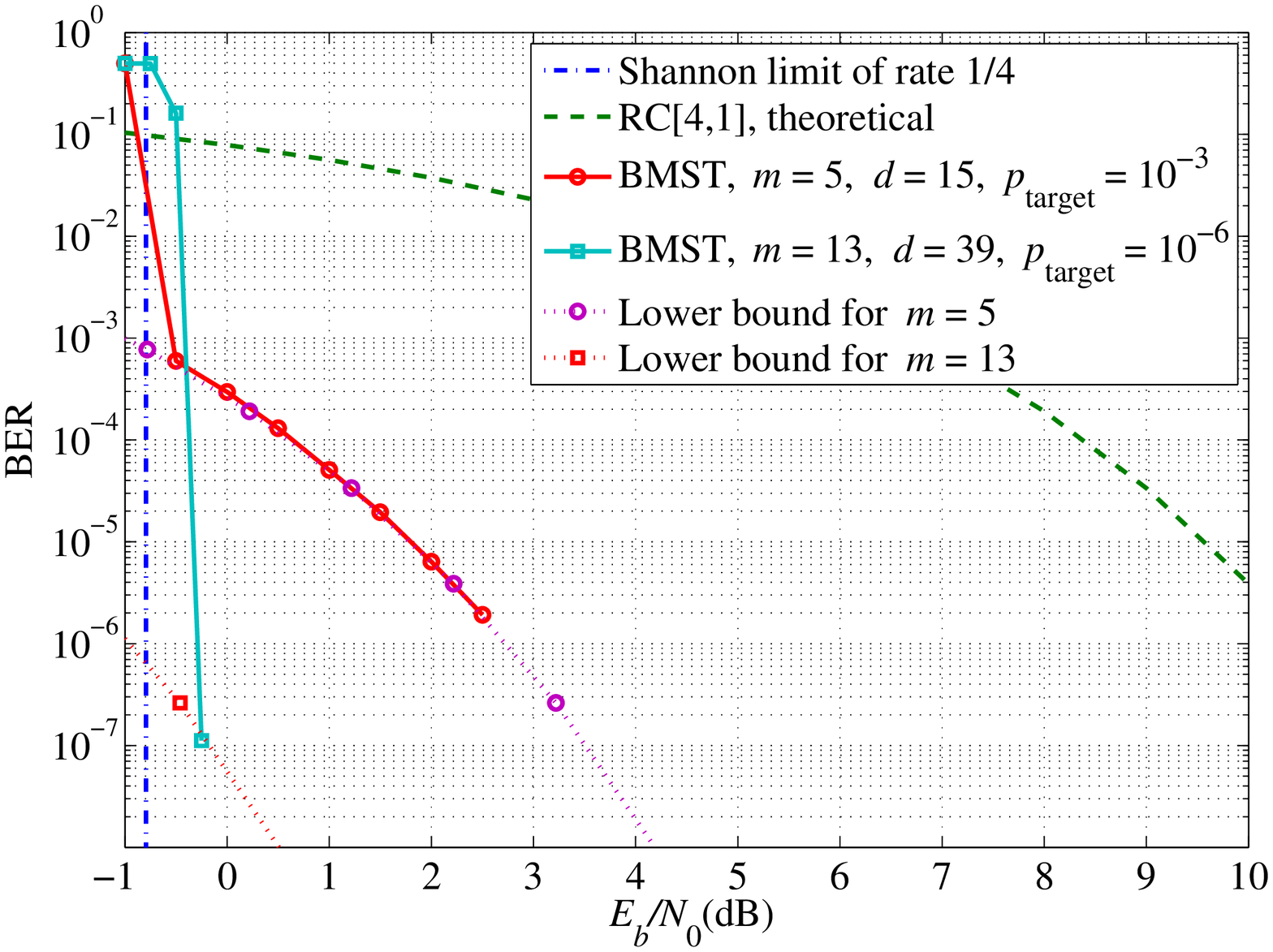}
    \caption{Performance of the BMST systems with the RC $[4,1]^{2500}$ as the basic code. The target BERs are $10^{-3}$ and $10^{-6}$. The systems encode $L=100000$ sub-blocks of data and decode with the SWD algorithm of a maximum iteration $I_{\max}=18$.}
    %and a stopping threshold $\delta_h=10^{-5}$, where encoding memories and the decoding delays are specified in the legend.}
   \label{fig:rc41m3}
\end{figure}

\begin{figure}[t]
   \centering
   \includegraphics[width=\figwidth]{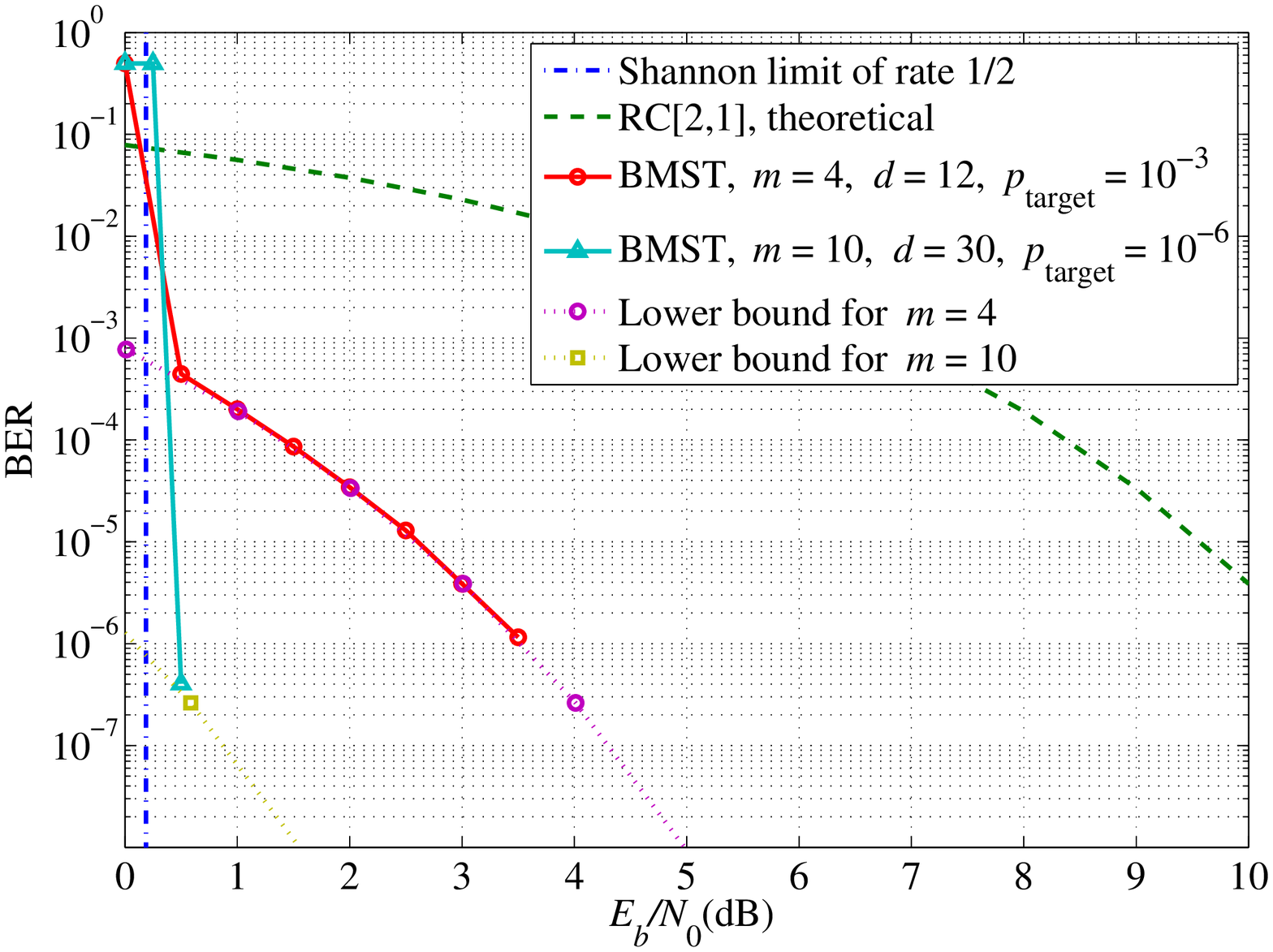}
    \caption{Performance of the BMST systems with the RC $[2,1]^{5000}$ as the basic code. The target BERs are $10^{-3}$ and $10^{-6}$. The systems encode $L=100000$ sub-blocks of data and decode with the SWD algorithm of a maximum iteration $I_{\max}=18$.}
    %and a stopping threshold $\delta_h=10^{-5}$, where encoding memories and the decoding delays are specified in the legend.}
   \label{fig:rc21m3}
\end{figure}

\begin{figure}[t]
   \centering
   \includegraphics[width=\figwidth]{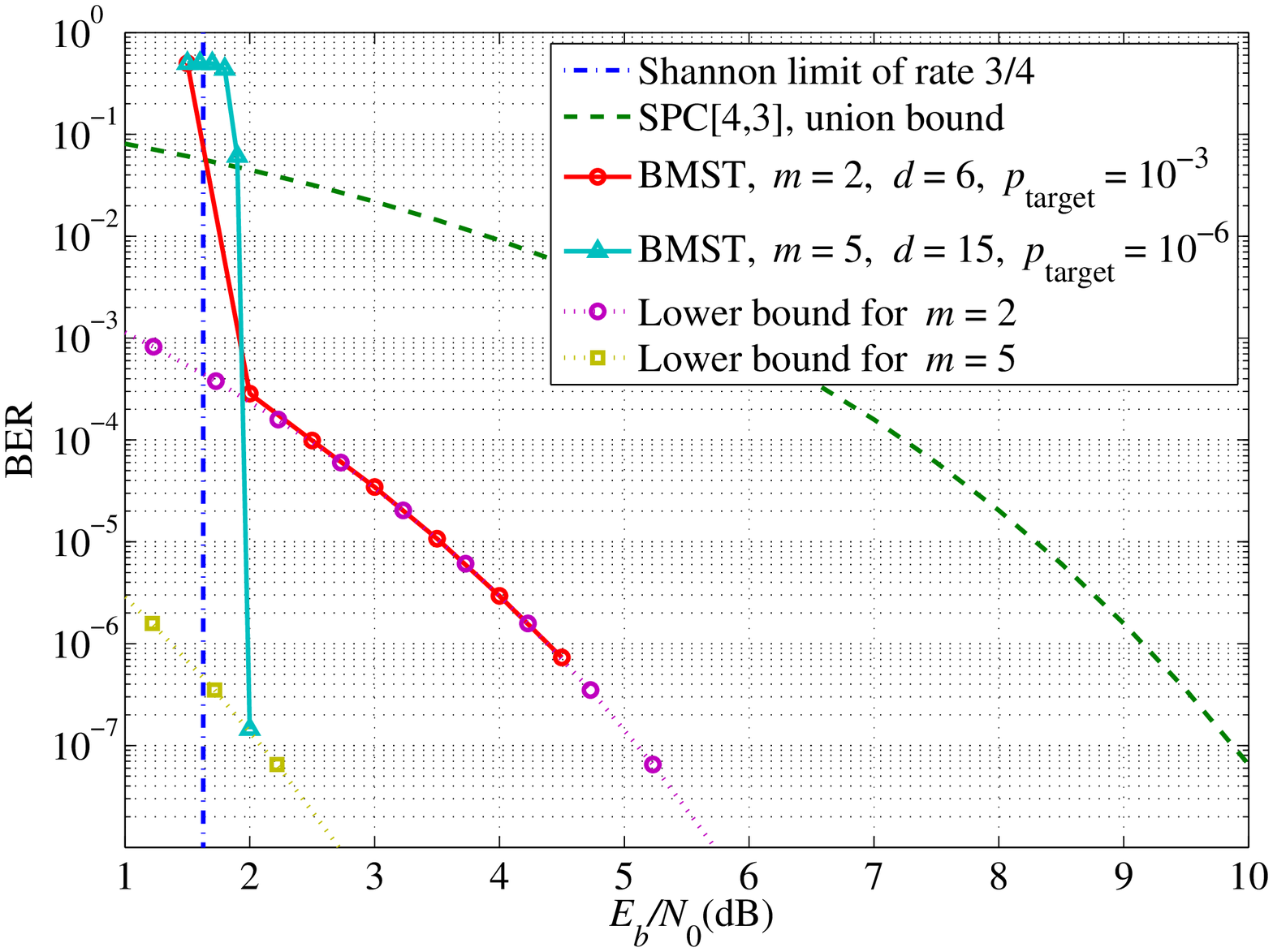}
    \caption{Performance of the BMST systems with the SPC $[4,3]^{2500}$ as the basic code. The target BERs are $10^{-3}$ and $10^{-6}$. The systems encode $L=100000$ sub-blocks of data and decode with the SWD algorithm of a maximum iteration $I_{\max}=18$.}
    %and a stopping threshold $\delta_h=10^{-5}$, where encoding memories and the decoding delays are specified in the legend.}
   \label{fig:spc43m3}
\end{figure}

\begin{figure}[t]
   \centering
   \includegraphics[width=\figwidth]{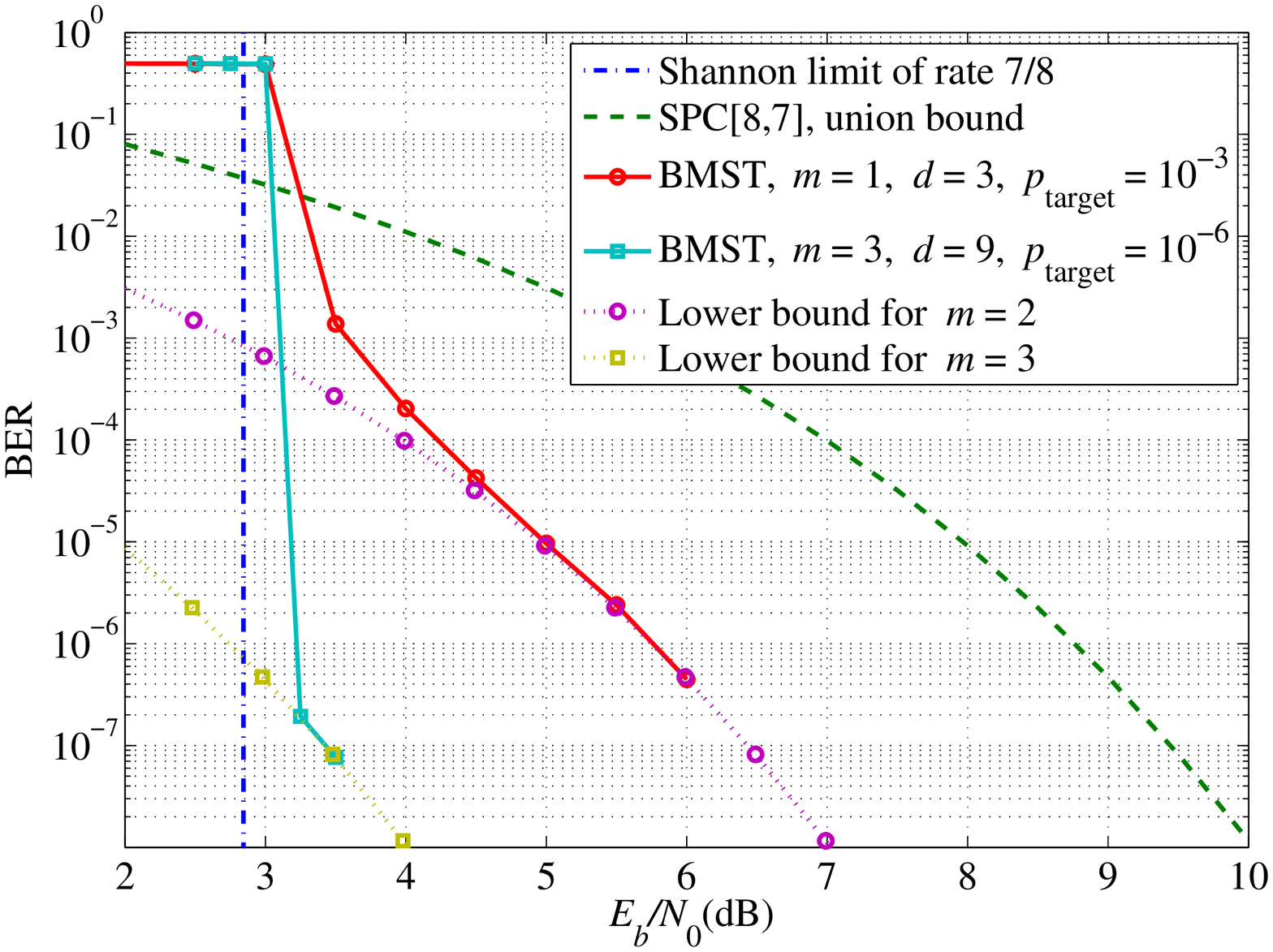}
    \caption{Performance of the BMST systems with the SPC $[8,7]^{1250}$ as the basic code. The target BERs are $10^{-3}$ and $10^{-6}$. The systems encode $L=100000$ sub-blocks of data and decode with the SWD algorithm of a maximum iteration $I_{\max}=18$.}
    %and a stopping threshold $\delta_h=10^{-5}$, where encoding memories and the decoding delays are specified in the legend.}
   \label{fig:spc87m3}
\end{figure}

We have conducted simulations for all constructed examples above with the target BERs $10^{-3}$ and $10^{-6}$.
Since all chosen basic codes have the form of Cartesian product, the brute-force MAP decoding algorithm based on Bayes' rule is implemented as the SISO decoding algorithm that is embedded in the iterative SWD algorithm.
In all simulations, we set $L = 100000$ for the encoder and a maximum iteration number $I_{\max} = 18$~(with
the entropy-based early stopping criterion~\cite{Ma04}) for the SWD algorithm. The decoding delay is set to be $d = 3m$.
Simulation results together with the lower bounds are presented in Figures~\ref{fig:rc81m3}-\ref{fig:spc87m3}.
From these numerical results, we have the following observations.
\begin{enumerate}
  \item The performance curves match well with the corresponding lower bounds, implying that the SWD algorithm is near optimal for the BMST in the low error rate region.
  \item All constructed BMST systems have achieved the respective target BERs within one dB away from the corresponding Shannon limits, implying that the proposed procedure is effective\footnote{Actually, this effectiveness has also been confirmed in the scenario where the high-order modulation is implemented over either AWGN channels or fast Rayeigh fading channels~\cite{Liang14}.}.
\end{enumerate}
%incorporated
%In the following, we present simulation performances of the SWD algorithm for the above constructed BMST systems at the target BERs $10^{-3}$ and $10^{-6}$. For all the presented constructions, all interleavers are uniformly random interleavers~(randomly generated but fixed). As we take the Cartesian product of some short block code as the basic code, the brute-force MAP decoding algorithm based on Bayes' rule is implemented as the SISO decoding algorithm for the SWD algorithm. In all simulations, we set $L=100000$ for the encoder and a maximum iteration $I_{\max}=18$~(with the entropy-based early-stopping criterion~\cite{Ma04}) for the SWD algorithm. {\color{red}Simulation results together with the lower bounds are presented in Fig.~\ref{fig:rc81m3} to Fig.~\ref{fig:spc87m3}.} For all simulated BMST systems, the performance curves of the SWD algorithm match well with the corresponding lower bounds {\color{red}at the low BER region.} In addition, each of the simulated BMST systems has a BER lower than the target BER within one dB away from the corresponding Shannon limits. In a word, the SWD algorithm is near-optimal for the BMST in the lower error rate region.

\section{The Two-phase Decoding for BMST}\label{sec:TwoPhaseDecoding}
We have illustrated by simulation both the effectiveness of the proposed procedure and the near-optimality of the SWD algorithm at BER around $10^{-5}$ or higher. An unavoidable question is that whether or not this procedure is applicable to the case when the target BER is extremely low. The difficulty lies in the fact that it is very time-consuming and even infeasible to verify this matchness by conventional software simulation in the extremely low error rate region.
{In this case, importance sampling, hardware emulation and/or other techniques~(see~\cite{Dolecek09} and the references therein) can be used to predict the error floor.
Here we propose a \emph{two-phase decoding}~(TPD) algorithm for the BMST system, whose performance is predictable.} For doing so, we propose a genie-aided decoder and a corresponding genie-aided bound, both of which are useful not only for developing the TPD algorithm but also for predicting the performance of the TPD algorithm.
%{\color{red}From~\cite{Ma13} as well as the previous section, we can see that the SWD algorithm delivers a performance curve around the BER of $10^{-5}$ that matches well with the genie-aided lower bound.} Although we conjecture that this is a general behavior~\cite{Ma13x}, it is very time-consuming to verify this matchness by simulation in the extremely low error rate region.

%\vpscaesection
%\section{Genie-aided Bounds}\label{sec:UpperBound}
\subsection{A Genie-aided Decoder}\label{subsec:genieAidedDecoder}
Let $\mathbfit{y} = \left( \mathbfit{y}^{(0)}, \cdots, \mathbfit{y}^{(L+m-1)} \right)$ be the whole received sequence. In~\cite{Ma13x}, a \emph{genie-aided decoder}, which computes $\Prob{\!U_j^{(t)}=u | \mathbfit{u}',\!\mathbfit{y}\!},u\in\mathbb{F}_2$ for all $j$ and $t$ by assuming that the transmitted data $\mathbfit{u}' = \left(\mathbfit{u}^{(0)}, \cdots, \mathbfit{u}^{(t-1)}, \mathbfit{u}^{(t+1)}, \cdots, \mathbfit{u}^{(L-1)}\right)$ are available, has been used to derive the lower bound, where $U_j^{(t)}$ is the random variable corresponding to $u_j^{(t)}$.
This is equivalent to assuming a genie who tells the decoder all but one the intermediate codewords $\mathbfit{v}' = \left(\mathbfit{v}^{(0)}, \cdots, \mathbfit{v}^{(t-1)}, \mathbfit{v}^{(t+1)}, \cdots, \mathbfit{v}^{(L-1)}\right)$. This is also equivalent to assuming that $\mathbfit{w}'$ $=$ $\left\{ \mathbfit{w}^{(t',i)}\right.$ $=$ $\mathbfit{v}^{(t')} \mathbfit{\varPi}_{i}:$ $\left.\!-m \!\leq \!t' \!\leq \!L\!+\!m\!-\!1, t' \!\neq \!t, \!0 \!\leq \!i \!\leq \!m \!\right\}$ are available at the decoder. In this section, this assumption is relaxed by assuming a genie who tells the decoder $\mathbfit{w}'$ but with each digit being flipped independently with a probability $p_{\rm genie}$.
In terms of messages over the normal graph~(see Fig.~\ref{fig:decoder} for reference), $\mathbfit{w}'$ are the messages to the node of type \fbox{+} from the nodes of type \fbox{$\Pi$} that connect to all but the $t$-th layers. At the $(t+i)$-th node of type \fbox{+} representing $\mathbfit{C}^{(t+i)}\left(0 \leq i \leq m\right)$, the genie observes a noisy version of $\mathbfit{w}'$ through a binary symmetric channel~(BSC) with crossover probability $p_{\rm genie}$, denoted by
$\tilde{\mathbfit{w}}' = \left\{ \tilde{\mathbfit{w}}^{(t',i)}: -m \!\leq\! t' \!\leq\! L\!+\!m\!-\!1, t' \neq t, 0 \leq i \leq m \right\}$.
%With these assumptions, the genie-aided decoder can compute $\Prob{\!U_j^{(t)}=u | \tilde{\mathbfit{w}}', \!\mathbfit{y}\!}\left(u\in\mathbb{F}_2\right)$ for all $j$ and make decisions.
%However, such a decoder is not convenient for performance analysis. Instead,
{Upon the observation of $\tilde{\mathbfit{w}}'$, the genie recovers ${\mathbfit{u}}^{(t)}$ by performing the following decoding algorithm. Notice that, since the parameter $p_{\rm genie}$ is not taken into account, this decoding algorithm is not optimal.}
\begin{algorithm}{\bf \em The Genie-aided Decoding for BMST}\label{alg:GenieAidedDecoding}

At time $t$, perform the following steps to recover ${\mathbfit{u}}^{(t)}$.
\begin{itemize}
  \item{{\bf Cancelation}:} \label{step:Cancelation}
    Remove the effect of $\tilde{\mathbfit{w}}'$ from $\mathbfit{y}^{(t)}, \cdots, \mathbfit{y}^{(t+m)}$ as if it were ${\mathbfit{w}}'$ by computing
    \begin{equation}
    \tilde{\mathbfit{c}}^{(t+i)} = \sum\limits_{\ell=0,\ell \neq i}^{m} \tilde{\mathbfit{w}}^{(t+i-\ell, \ell)}
    \end{equation}
    and
    \begin{equation}
    \tilde{\mathbfit{y}}^{(t+i)} =  \left(-1\right)^{\tilde{\mathbfit{c}}^{(t+i)}} {\scriptstyle \bigodot} \mathbfit{y}^{(t+i)}
    \end{equation}
    for $0 \leq i \leq m$, where ${\scriptstyle \bigodot}$ denotes the component-wise multiplication between two vectors.
  \item{{\bf Minimization}:} \label{step:Minimization}
    Find $\hat{\mathbfit{u}}^{(t)}$~(equivalently $\hat{\mathbfit{v}}^{(t)}$) that minimizes $\sum_{i=0}^{m}\left\| \tilde{\mathbfit{y}}^{(t+i)}\!-\!\left(-1\right)^{\hat{\mathbfit{v}}^{(t)}\mathbfit{\varPi}_{i}}\right\|^2$, where $\|\cdot\|^2$ represents the squared Euclidean norm.
  \item{{\bf Output}:} \label{step:decoding}
    Output $\hat{\mathbfit{u}}^{(t)}$ as the decoding result.
\end{itemize}
\end{algorithm}
\textbf{Remarks.} Note that $\tilde{\mathbfit{c}}^{(t+i)}$ is irrelevant to $\mathbfit{w}^{(t,i)}$, that is, $\tilde{\mathbfit{c}}^{(t+i)}$ is irrelevant to $\mathbfit{v}^{(t)}$ for $0\!\leq\!i\!\leq\!m$. Also note that Step~{\bf Minimization} is implementable when the basic code is
%either a convolutional code with a short constraint length or a Cartesian product of a short block code. For example, if the basic code is
$\mathscr{C}\left[N,K\right]^B$, a Cartesian product of a short block code. In this case, the minimization is indeed $B$ separate and independent minimizations, each of which has complexity at most $O(2^K)$. For the same reason, we can safely assume that $B = 1$ in the following for the BER performance analysis.

\vpscaesection
\subsection{Upper Bound for the Genie-aided Decoder}\label{subsec:GenieAidedBound}
%To derive the genie-aided bound for the genie-aided decoder, $\tilde{\mathbfit{w}}'$ should satisfy the following two assumptions\footnote{Actually, these assumptions are needed for deriving the genie-aided bound and irrelevant to the implementation of the genie-aided decoding~(GAD) algorithm.}:
To derive the genie-aided bound for the genie-aided decoder, we need the following two assumptions.
Notice that these two assumptions are required only for performance analysis but not for implementing the genie-aided decoding~(GAD) algorithm.
\begin{itemize}
  \item{\bf Assumption 1.} The components of $\tilde{\mathbfit{w}}'$ are statistically independent.
  \item{\bf Assumption 2.} Each component $\tilde{{w}}'^{(t,i)}_{j}$ of $\tilde{\mathbfit{w}}'$ is statistically independent of the corresponding received signal ${y}^{(t+i)}_{j}$.
\end{itemize}

With the above two assumptions, we can see that each bit $v^{(t)}_j$ is transmitted $m+1$ times through a BI-AWGNC with a flipping error. The flipping error occurs whenever there are an odd number of errors among the $m$ summands that contribute to $\tilde{c}^{(t+i)}_j$. Hence, the probability of a coded bit being flipped is given by~\cite{Gallager63}
\begin{eqnarray}\label{eq:bscPe}
    \bscPe &=& \sum\limits_{\sumLayerBit \mbox{ is odd}} \binom{m}{\sumLayerBit} {\phaseOnePe}^{\sumLayerBit} \left(1-{\phaseOnePe}\right)^{m-\sumLayerBit} \nonumber \\
     &=& \frac{1-\left(1-2\phaseOnePe\right)^m}{2}.
\end{eqnarray}
{Assume that $\mathbfit{u}^{(t)} = \mathbf{0}$ is transmitted but $\hat{\mathbfit{u}}^{(t)} \neq \mathbf{0}$ is  the decoding output, whose corresponding codeword $\hat{\mathbfit{v}}^{(t)}$ has Hamming weight $h$.} Then the pairwise error probability~(PEP), denoted by $\Prob{\mathbfit{u}^{(t)} \rightarrow \hat{\mathbfit{u}}^{(t)}}$, is the same as the probability
\begin{align}\label{eq:PEP}
    \Prob{\! \sum_{i=0}^{m} \left\| \tilde{\mathbfit{y}}^{(t+i)} \!-\! \left(\!-\!1\!\right)^{\mathbfit{w}^{(t,i)}}  \right\|^2 \!\geq\! \sum_{i=0}^{m} \left\| \tilde{\mathbfit{y}}^{(t+i)} \!-\! \left(\!-\!1\!\right)^{\hat{\mathbfit{w}}^{(t,i)}} \right\|^2 \!} \nonumber \\
    = \Prob{ \sum_{i=0}^{m}\sum_{j=0}^{N-1} \tilde{{y}}^{(t+i)}_{j} \left( 1 - \left(-1\right)^{\hat{{w}}^{(t,i)}_{j}} \right) \leq 0 },
\end{align}
where $\mathbfit{w}^{(t, i)} = \mathbfit{v}^{(t)} \mathbfit{\varPi}_i$, $\hat{\mathbfit{w}}^{(t,i)}=\hat{\mathbfit{v}}^{(t)}\mathbfit{\varPi}_{i}$ and $\tilde{{y}}^{(t,i)}_{j}$ and $\hat{{w}}^{(t,i)}_{j}$ are the $j$-th components of $\tilde{\mathbfit{y}}^{(t+i)}$ and $\hat{\mathbfit{w}}^{(t,i)}$, respectively. It is not surprising that the PEP depends only on the Hamming weight $W_{\rm H}\left( ( \hat{\mathbfit{w}}^{(t,0)}, \cdots, \hat{\mathbfit{w}}^{(t,m)}) \right) = \left(m+1\right)h$. Since each $\tilde{y}^{(t)}_{j}$ is distributed according to $\mathcal{N}\left(-1, \sigma^2\right)$ with probability $\bscPe$ and $\mathcal{N}\left(+1, \sigma^2\right)$ with probability $1-\bscPe$, we have
\begin{align}\label{eq:ProbSumY}
    \Prob{\mathbfit{u}^{(t)} \rightarrow \hat{\mathbfit{u}}^{(t)}} \!=\! \sum_{\errBitNum=0}^{\left(m\!+\!1\right)\nWeightIdx}\! \binom{\!\left(m\!+\!1\right)\!\nWeightIdx}{\!\errBitNum} \cdot~~~~~~~~~~ \nonumber \\
    \!\bscPe^{\errBitNum}\!\left(1\!-\!\bscPe\right)^{\left(m\!+\!1\right)\nWeightIdx-\errBitNum} \!\Qfun{\!\frac{\left(m\!+\!1\right)\nWeightIdx-2\errBitNum}{\sqrt{\left(m\!+\!1\right)\nWeightIdx}\sigma}\!}.
\end{align}
Using the union bound, the genie-aided bound is given by
\begin{equation}\label{eq:UnionBoundEqSys}
  f_{\rm genie}(\gamma_b) \leq \sum\limits_{\kWeightIdx=1}^{\CarK} \sum\limits_{\nWeightIdx=1}^{\CarN} \frac{\kWeightIdx}{\CarK} A_{\kWeightIdx,\nWeightIdx} \Prob{\mathbfit{u}^{(t)} \rightarrow \hat{\mathbfit{u}}^{(t)}},
\end{equation}
where $\gamma_b=10\log_{10}\left(\frac{1}{2\sigma^{2} \cdot R}\right)$ and $R$ is the code rate of the basic code.

\begin{example}[Genie-aided bounds]%[Example~\ref{ex:BMSTwithRC} continued]
\label{ex:UpperBound}
Consider the BMST system with the RC $[2, 1]^{5000}$ of an encoding memory $m=30$, as determined in Example~\ref{ex:BMSTwithRC}. Hence, the genie-aided system is equivalent to a system that transmit an RC $[2,1]$ codeword $31$ times through a BI-AWGNC with a flipping error $p_{\rm flip}$.
%We take the target BER of $10^{-15}$ with a required encoding memory $m=30$ for example. The bound computed by \Ref{eq:UnionBoundEqSys} is also the performance function of the genie-aided decoder.
Substituting \Ref{eq:IOWEF_RC}, \Ref{eq:bscPe} and \Ref{eq:ProbSumY} into \Ref{eq:UnionBoundEqSys}, we have
\begin{align}\label{eq:UpperBoundwithRC}
  f_{\rm genie}(\gamma_b) = \sum_{ \errBitNum = 0}^{62} \binom{62}{\errBitNum} \cdot~~~~~~~~~~~~~~~~~~~~~~~~~~~~~~~~~~~~~~ \nonumber \\
   \!\frac{\left[\!1\!-\!\left(\!1\!-\!2\phaseOnePe\!\right)^{30}\right]^{\errBitNum}
  \!\left[\!1\!+\!\left(\!1\!-\!2\phaseOnePe\!\right)^{30}\right]^{62\!-\!\errBitNum}}{2^{62}} \cdot \Qfun{\!\frac{62\!-\!2\errBitNum}{\sqrt{62}\sigma}},
\end{align}
where $\sigma$ can be determined by $\gamma_b=10\log_{10}\left(\frac{1}{2\sigma^{2}\cdot (1/2)}\right)$.
The bounds for the genie-aided decoder of this BMST system with different $\phaseOnePe$ are plotted in Fig.~\ref{fig:BMSTwithRC}. We can see that the bound gets lower as $\phaseOnePe$ becomes smaller. For this example, by setting $\phaseOnePe=0$, the proposed genie-aided bound is reduced to the genie-aided lower bound~\cite{Ma13,Ma13x}. With $\phaseOnePe=10^{-5}$, the genie-aided bound is very close to the lower bound.
\begin{figure}[t]
   \centering
   \includegraphics[width=\figwidth]{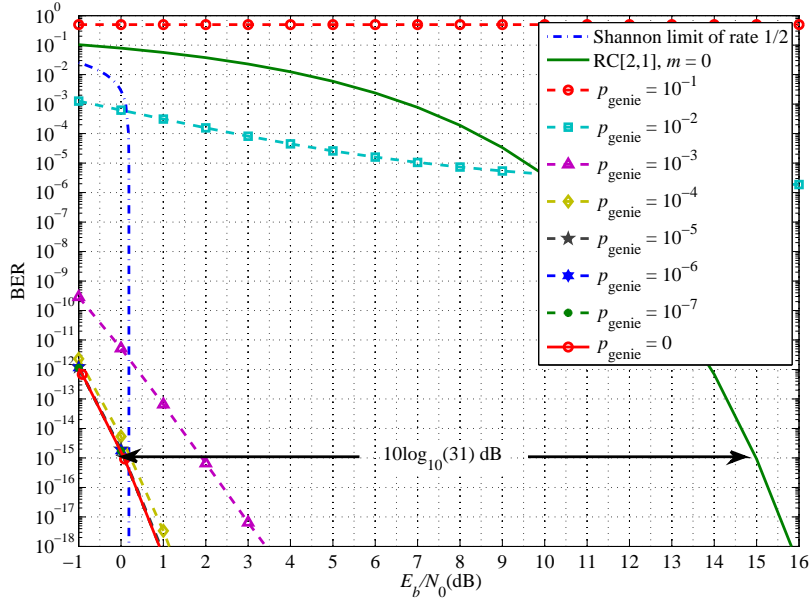}
    \caption{The genie-aided bounds with different $\phaseOnePe$ for the BMST system with an encoding memory $m=30$ using RC~$[2,1]^{B}$ as the basic code. The performance curve for the basic code~($m=0$) is also plotted here.}
   \label{fig:BMSTwithRC}\vpscaefigure
\end{figure}
\end{example}
\textbf{Remark.}~Notice that this genie-aided bound will not drop down to zero even when $E_b/N_0$ goes to infinity. As $E_b/N_0$ goes to infinity, the channel model for the genie-aided bound is changed into a BSC, where the BER is determined by $p_{\rm genie}$.

\subsection{Two-phase Decoding}
%It has been demonstrated by simulation in~\cite{Ma13} that the SWD algorithm delivers a performance curve around the BER of $10^{-5}$ that matches well with the genie-aided lower bound. Although we conjecture that this is a general behavior~\cite{Ma13x}, it is very time-consuming to verify this matchness by simulation in the extremely low error rate region. Hence we propose here a two-phase decoding algorithm for the BMST system, whose performance can be predictable.
{As seen from Example~\ref{ex:UpperBound}, if we could find a ``genie" who tells us ${\mathbfit{w}}'$ with an error probability $p_{\rm genie} \approx 10^{-5}$, we can perform the GAD algorithm. If this is the case, the performance in the extremely low error rate region can be predicted with the help of the genie-aided bound.}
{Motivated by this, we propose a two-phase decoding~(TPD) algorithm, where the phase-I decoding performs the original SWD algorithm, which serves as a ``genie". The phase-II decoding performs the GAD algorithm, which attempts to clean up the residual errors once when the phase-I decoding lowers the BER down to around $10^{-5}$.}
%So far, the genie-aided bound has be derived. However, to apply the genie-aided bound to the prediction of the performance of the BMST in the extremely low error rate region, we should find a ``genie". As a result, we propose a two-phase decoding~(TPD) algorithm, where the phase-I decoding performs the original SWD algorithm, which serves as a ``genie". The phase-II decoding performs the GAD algorithm, which ``cleans up" the residual errors once when the phase-I decoding lowers the BER down to around $10^{-5}$.

%In~\cite{Ma13x}, the SWD algorithm matches well with the lower bound at the BER of around $10^{-5}$. However, there is no effective method to analyze the performance of the SWD algorithm in the extremely low error rate region, while simulating the performance is time-consuming. Hence, we propose a two-phase decoding algorithm for the BMST, of which the performance is predictable. For the TPD algorithm, the original SWD algorithm is used as the phase-I decoding, while the genie-aided decoding~(GAD) algorithm is used as the phase-II decoding to remove the residual~(rare) errors in the outputs of the phase-I decoding, where the outputs of the phase-I decoding are treated as told by a genie.
Upon receiving $\mathbfit{y}^{(t)}$, the TPD algorithm of a decoding delay $d$ for a BMST system of memory $m$ is performed to recover $\mathbfit{u}^{(t-d-m)}$ over the corresponding normal graph as follows.
\begin{algorithm}{\bf \em Two-phase Decoding of BMST}\label{alg:TwoPhase}
\begin{itemize}
  \item {\bf{Initialization}:} For $t\!=\!0$, $1$, $\cdots$, $d\!-\!1$, compute $\Prob{ C^{(t)}_{j}\!=\!u | {y}^{(t)}_{j}\!},u\!\in\!\mathbb{F}_2$ and initialize the messages associated with $\mathbfit{C}^{(t)}$ by $\Prob{ C^{(t)}_{j}\!=\!u | {y}^{(t)}_{j}\!},u\!\in\!\mathbb{F}_2 \left(0\!\leq\!j\!\leq\!n\!-\!1\right)$. All messages over the other edges within and connecting to the $t$-th layer are initialized as uniformly distributed variables.

  \item {\bf{Two-phase decoding}:} For $t\!=\!d$, $d+1$, $\cdots$, $L\!+\!m\!-\!1$,

        {{\underline{Phase I}}:} \label{step:Phase1}
           Perform the SWD algorithm with a decoding delay $d\!>\!0$ and a maximum iteration number $I_{\max}\!>\!0$ using the entropy-based early stopping criterion~\cite{Ma13x}. Denote the decoding outputs as $\tilde{\mathbfit{w}}^{(t-d,0)}, \cdots, \tilde{\mathbfit{w}}^{(t-d,m)}$, which are the hard decisions on the {\em extrinsic} messages to the node of type \fbox{+} from the nodes of type \fbox{$\Pi$} that connect to the $(t-d)$-th layer.

        {{\underline{Phase II}}:} \label{step:Phase2}
           If $t\!\geq\!d\!+\!m$, perform the GAD algorithm to recover ${\mathbfit{u}}^{(t-d-m)}$ by treating the decoding outputs $\left\{ \tilde{\mathbfit{w}}^{(t',i)} : t\!-\!d\!-\!2m \!\leq\! t' \!\leq\! t \!-\! d, t' \neq t\!-\!d\!-\!m, 0 \!\leq\! i \!\leq\! m \right\}$ of \underline{Phase~I} as what were told by a genie. Denote the decoding output as $\hat{\mathbfit{u}}^{(t-d-m)}$.

  \item {\bf{Output}:} Output $\hat{\mathbfit{u}}^{(t)}\left(0 \leq t \leq L-1\right)$ as the decoding results.
\end{itemize}
\end{algorithm}
\vpscaesection
\subsection{Performance Analysis of the TPD algortihm}
Let $p_{\rm I}$ be the average BER of $\tilde{\mathbfit{w}}^{(t,i)}\left(0 \!\leq\! i \!\leq\! m, 0 \!\leq\! t \!\leq\! L\!-\!1\right)$ and $p_{\rm II}$ be the average BER of $\hat{\mathbfit{u}}^{(t)}\left(0 \leq t \leq L-1\right)$. Note that $p_{\rm I}$ is not identical in concept to the BER of the original SWD algorithm. As widely accepted, if $p_{\rm I} \approx 10^{-5}$, it can be estimated reliably by simulation.
{However, conventional software simulation becomes helpless to estimate $p_{\rm II}$ if it is around $10^{-15}$. The question is, can we predict $p_{\rm II}$ using $p_{\rm I}$ with the help of the genie-aided bound?}
%However, in the case of $p_{\rm I} \approx 10^{-5}$, the phase-II decoding might~(very likely for large $m$) correct all residual errors. If so, we need apply the genie-aided bound to $p_{II}$ for performance prediction.

To answer this question, we need to check if the outputs from the phase-I decoding satisfy the two assumptions as presented in Section~\ref{subsec:GenieAidedBound}.
Fortunately, in the case when $n \gg m$~(namely $B$ being large enough) and the $m+1$ interleavers are generated independently, these two assumptions hold with high probability.
Intuitively, these assumptions, which are similar to what have been widely used in the performance analysis of LDPC codes~\cite{Richardson01}, are reasonable due to the existence of the random interleavers~(of large size) and the features of the extrinsic messages.
{However, for a rigorous proof~(omitted here), we need to define a message flow neighborhood, similar to~\cite{Kavcic03}, by replacing the trellis constraint~(representing the intersymbol interference channel) in~\cite[Fig.~8 and Fig.~9]{Kavcic03} with the constraint induced by the basic code.}
\begin{example}[Validity of the genie-aided bounds]%[Example~\ref{ex:BMSTwithRC} continued]
\label{ex:UpperBoundValidity}
To confirm the upper bound, we take as an example the BMST system with the RC $[2,1]^{5000}$ of $m=8$ constructed in Example~\ref{ex:BMSTwithRC}.
%We take the target BER of $10^{-5}$ as an example to show the validity of the upper bound.
In this case, both $p_{\rm I}$ and $p_{\rm II}$ can be estimated reliably by simulation. The simulation results for the BMST system with the RC $[2,1]^{5000}$ as the basic code are shown in Fig.~\ref{fig:rc212m8}. In both Fig.~\ref{fig:rc212m8} and Fig.~\ref{fig:rc212m30}, the performance curves of the SWD algorithm corresponding to the BER of $\mathbfit{u}^{(t)}$ are denoted as ``SWD", while the performance curves of the TPD algorithm corresponding to $p_{\rm I}$ and $p_{\rm II}$ are denoted as ``TPD, $p_{\rm I}$" and ``TPD, $p_{\rm II}$", respectively. We can see that the simulation results of the TPD algorithm match well with the upper bound  in the low BER region.
\begin{figure}[t]
   \centering
   \includegraphics[width=\figwidth]{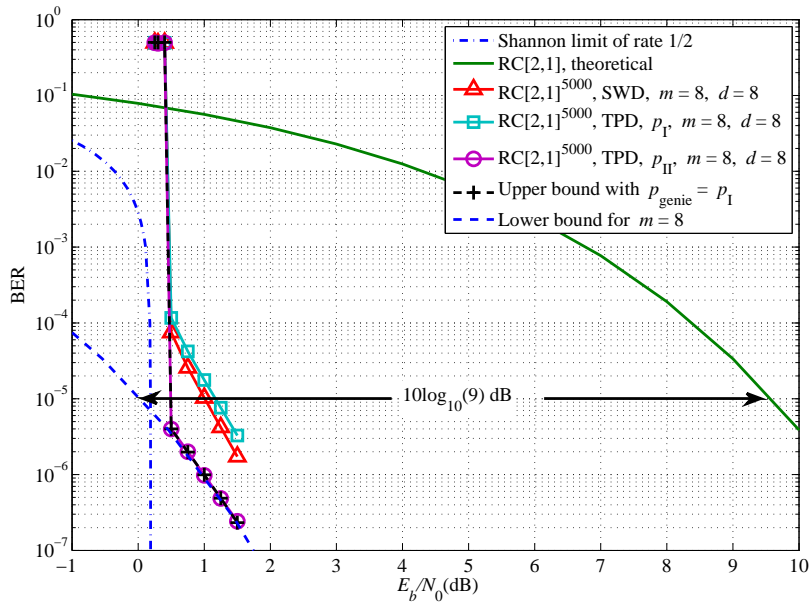}
    \caption{Performance of the BMST system with the RC $[2,1]^{5000}$ as the basic code. The target BER is $10^{-5}$. The system encodes $L=100000$ sub-blocks of data with the encoding memory $m=8$ and decodes with a decoding delay $d=8$ and a maximum iteration $I_{\max}=18$.}
   \label{fig:rc212m8}\vpscaefigure
\end{figure}
%As simulations can be done for both the SWD algorithm and the TPD algorithm at a BER of around $10^{-5}$, we take the target BER of $10^{-5}$ as an example to show the validity of the upper bound. Simulation results for the BMST system with the RC~$[2,1]^{5000}$ as the basic code are shown in Fig.~\ref{fig:rc212m8}. The curve for the SWD algorithm corresponding to the BER of $\mathbfit{u}^{(t)}$ and the curves of the TPD algorithm corresponding to $p_{\rm I}$ and $p_{\rm II}$ are specified in the figure. In this example, the upper bound computed from $p_{\rm I}$ of the respective $E_b/N_0$ is also the performance of the phase-II decoding. We can see that the BER performance of the TPD algorithm matches well with the upper bound in the low BER region.
%We can also see that which means that simulating the performance of the phase-I decoding down to a BER of about $10^{-5}$ is enough to predict the performance of the TPD algorithm in the extremely low BER region.
%\begin{table}[t]
%\caption{\label{tab:UpperBound}}
%\begin{tabular}{l|ccc}
%  \hline
%  \hline
%  % after \\: \hline or \cline{col1-col2} \cline{col3-col4} ...
%  $E_b/N_0$ (dB) & $0.5$ & $1.0$ & $1.5$ \\
%  \hline
%  $p_{\rm I}$  & 0.0000705007 & 0.0000105285 & 0.0000016725 \\
%  $p_{\rm II}$ & 0.0000039002 & 0.0000010946 & 0.0000002438 \\
%  Bound on $p_{\rm II}$ & 0.0000037995 & 0.0000009839 & 0.0000002311 \\
%  \hline
%  \hline
%\end{tabular}
%\end{table}
\end{example}

\begin{example}[A construction with a target BER $10^{-15}$]%[aaaaaaa]
\label{ex:NumericalResults}
Consider the BMST system using the RC $[2,1]^{5000}$ constructed with a target BER $10^{-15}$ in Example~\ref{ex:BMSTwithRC}.
%For this BMST system, we present the simulation performance of the SWD algorithm and predict the performance of the TPD algorithm.
%In both the SWD algorithm and the phase-I decoding, the brute-force MAP decoding algorithm based on Bayes' rule is implemented as the SISO decoding algorithm for the basic code.
In the simulations, we set $L=100000$ for the encoder and a maximum iteration number $I_{\max}=18$ and a decoding delay $d = 2m = 60$ for both the SWD algorithm and the TPD algorithm. Simulation results for the SWD algorithm together with the lower bound and the genie-aided upper bound are shown in Fig.~\ref{fig:rc212m30}.
As we can see, at $E_b/N_0=0.5$~dB, the BER of the phase-I decoding $p_{\rm I} = 7.0 \times 10^{-6}$. Although no error bits are collected at $E_b/N_0=0.5$~dB for the phase-II decoding during our simulations, the performance of the TPD algorithm can be predicted as $p_{\rm II} = 4.2 \times 10^{-17}$ by substituting $p_{\rm genie} = p_{\rm I} = 7.0 \times 10^{-6}$ into the genie-aided bound~(\ref{eq:UpperBoundwithRC}). It is then safe to conclude that the constructed BMST system can approach the Shannon limit within one dB at the BER of $10^{-15}$.

%As we can see, at $E_b/N_0 = 0.5$~dB, the BER of the phase-I decoding $p_{\rm I} = 7.0 \times 10^{-6}$, while no error bit is collected for the phase-II decoding during our simulation. Hence, it can be predicted that the TPD algorithm has a performance $p_{\rm II} = 4.2 \times 10^{-17}$ according to the genie-aided bound computed using~(\ref{eq:UpperBoundwithRC}) by setting $p_{\rm genie} = p_{\rm I} = 7.0 \times 10^{-6}$, implying that the constructed BMST system can approach the Shannon limit within one dB at a BER lower than $10^{-15}$.

%we have simulated 4.134 ¡Á 106 sub-blocks of data (3.7206 ¡Á 1010 information bits in total) for the TPD algorithm with a decoding delay d = 16, where no error bit is collected for both the phase-I decoding and the phase-II decoding.
\begin{figure}[t]
   \centering
   \includegraphics[width=\figwidth]{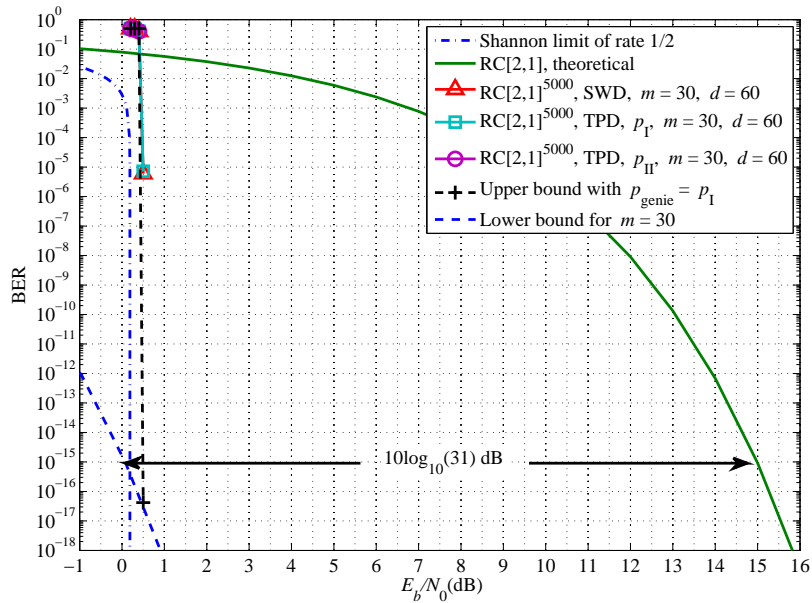}
    \caption{Performance of the BMST system with the RC $[2,1]^{5000}$ as the basic code. The target BER is $10^{-15}$. The system encodes $L=100000$ sub-blocks of data with the encoding memory $m=30$ and decodes with a decoding delay $d=60$ and a maximum iteration $I_{\max}=18$.}
   \label{fig:rc212m30}\vpscaefigure
\end{figure}
\end{example}

\section{Conclusions}\label{sec:Conclusion}
In this paper, we have presented a general procedure for constructing block Markov superposition transmission~(BMST) systems to approach the corresponding Shannon limits. The procedure is shown by simulation to be effective for a variety of code rates and a variety of target BERs. We have also proposed a two-phase decoding~(TPD) algorithm for the BMST, whose BER performance is predictable given the BER performance of the outputs of the phase-I decoding. A BMST with memory $m=30$ constructed using the Cartesian product of the repetition code~(RC) $[2,1]^{5000}$ can approach the Shannon limit within one dB at the BER of $10^{-15}$.

\section*{Acknowledgment}
The authors are grateful to Mr. Kechao Huang for useful discussions. The authors are also grateful to Prof. Costello for his suggestion to make the connection of BMST with spatial coupling.
%This work is supported by the 973 Program (No.2012CB316100) and the NSF (No.61172082) of China.
%The authors are grateful to Mr. Kechao Huang for useful discussions and Prof. Daniel J. Costello, Jr. for suggesting that we connect block Markov superposition transmission~(BMST) with spatial coupling.

%% References:
%% We recommend the usage of BibTeX:
%%

%\bibliographystyle{IEEEtran}
%\bibliography{IEEEabrv,liang}

\begin{thebibliography}{10}
\providecommand{\url}[1]{#1}
\csname url@samestyle\endcsname
\providecommand{\newblock}{\relax}
\providecommand{\bibinfo}[2]{#2}
\providecommand{\BIBentrySTDinterwordspacing}{\spaceskip=0pt\relax}
\providecommand{\BIBentryALTinterwordstretchfactor}{4}
\providecommand{\BIBentryALTinterwordspacing}{\spaceskip=\fontdimen2\font plus
\BIBentryALTinterwordstretchfactor\fontdimen3\font minus
  \fontdimen4\font\relax}
\providecommand{\BIBforeignlanguage}[2]{{%
\expandafter\ifx\csname l@#1\endcsname\relax
\typeout{** WARNING: IEEEtran.bst: No hyphenation pattern has been}%
\typeout{** loaded for the language `#1'. Using the pattern for}%
\typeout{** the default language instead.}%
\else
\language=\csname l@#1\endcsname
\fi
#2}}
\providecommand{\BIBdecl}{\relax}
\BIBdecl

\bibitem{Berrou93}
C.~Berrou, A.~Glavieux, and P.~Thitimajshima, ``{Near Shannon limit
  error-correcting coding and decoding: Turbo codes},'' in \emph{Proc. Int.
  Conf. Commun.}, Geneva, Switzerland, May 1993, pp. 1064--1070.

\bibitem{Costello07}
D.~J. Costello, Jr. and G.~D. Forney, Jr., ``{Channel coding: The road to
  channel capacity},'' \emph{Proc. IEEE}, vol.~95, pp. 1150--1177, June 2007.

\bibitem{Marchant90}
A.~B. Marchant, \emph{{Optical Recording: A Technical Overview}}.\hskip 1em
  plus 0.5em minus 0.4em\relax New York: Addison-Wesley, 1990.

\bibitem{Sripimanwat05}
K.~Sripimanwat~(Ed.), \emph{{Turbo Code Applications: A Journey from a Paper to
  Realization}}.\hskip 1em plus 0.5em minus 0.4em\relax Dordrecht, The
  Netherlands: Springer, 2005.

\bibitem{Costello09}
\BIBentryALTinterwordspacing
D.~J. Costello, Jr., ``{The genesis of coding theory},'' in \emph{The Second
  Annual North American School of Information Theory}, Northwestern University,
  IL, Aug. 2009. [Online]. Available:
  \url{http://www.itsoc.org/conferences/past-schools/na-school-2009/material}
\BIBentrySTDinterwordspacing

\bibitem{Gallager63}
R.~G. Gallager, \emph{{Low-Density Parity-Check Codes}}.\hskip 1em plus 0.5em
  minus 0.4em\relax Cambridge, MA: MIT Press, 1963.

\bibitem{Richardson01}
T.~J. Richardson, M.~A. Shokrollahi, and R.~L. Urbanke, ``Design of
  capacity-approaching irregular low-density parity-check codes,'' \emph{IEEE
  Trans. Inf. Theory}, vol.~47, pp. 619--637, Feb. 2001.

\bibitem{Zhang10}
L.~Zhang, Q.~Huang, S.~Lin, K.~Abdel-Ghaffar, and I.~F. Blake, ``{Quasi-cyclic
  LDPC codes: An algebraic construction, rank analysis, and codes on Latin
  squares},'' \emph{IEEE Trans. Commun.}, vol.~58, pp. 3126--3139, Nov. 2010.

\bibitem{Huang12}
Q.~Huang, Q.~Diao, S.~Lin, and K.~Abdel-Ghaffar, ``{Cyclic and quasi-cyclic
  LDPC codes on constrained parity-check matrices and their trapping sets},''
  \emph{IEEE Trans. Inf. Theory}, vol.~58, pp. 2648--2671, May 2012.

\bibitem{Ma13}
X.~Ma, C.~Liang, K.~Huang, and Q.~Zhuang, ``{Obtaining extra coding gain for
  short codes by block Markov superposition transmission},'' in \emph{Proc.
  IEEE Int. Symp. Inf. Theory}, Istanbul, Turkey, July 2013, pp. 2054--2058.

\bibitem{Ma13x}
\BIBentryALTinterwordspacing
------, ``{Block Markov superposition transmission: Construction of big
  convolutional codes from short codes},'' submitted to {\it IEEE Trans. Inf.
  Theory}. [Online]. Available: \url{http://arxiv.org/abs/1308.4809}
\BIBentrySTDinterwordspacing

\bibitem{Felstrom99}
A.~J. Felstr\"{o}m and K.~S. Zigangirov, ``{Time-varying periodic convolutional
  codes with low-density parity-check matrix},'' \emph{IEEE Trans. Inf.
  Theory}, vol.~45, pp. 2181--2191, Sept. 1999.

\bibitem{Tanner04}
R.~M. Tanner, D.~Sridhara, A.~Sridharan, T.~E. Fuja, and D.~J. Costello, Jr.,
  ``{LDPC block and convolutional codes based on circulant matrices},''
  \emph{IEEE Trans. Inf. Theory}, vol.~50, pp. 2966--2984, Dec. 2004.

\bibitem{Lentmaier10}
M.~Lentmaier, A.~Sridharan, D.~J. Costello, Jr, and K.~Zigangirov, ``{Iterative
  decoding threshold analysis for LDPC convolutional codes},'' \emph{IEEE
  Trans. Inf. Theory}, vol.~56, pp. 5274--5289, Oct. 2010.

\bibitem{Pusane11}
A.~E. Pusane, R.~Smarandache, P.~O. Vontobel, and D.~J. Costello, Jr.,
  ``{Deriving good LDPC convolutional codes from LDPC block codes},''
  \emph{IEEE Trans. Inf. Theory}, vol.~57, pp. 835--857, Feb. 2011.

\bibitem{Kudekar11}
S.~Kudekar, T.~J. Richardson, and R.~L. Urbanke, ``{Threshold saturation via
  spatial coupling: Why convolutional LDPC ensembles perform so well over the
  BEC},'' \emph{IEEE Trans. Inf. Theory}, vol.~57, pp. 803--834, Feb. 2011.

\bibitem{Cheng96}
J.-F. Cheng, ``Hyperimposed convolutional codes,'' in \emph{Proc. Int. Conf.
  Commun.}, vol.~2, Dallas, TX, June 1996, pp. 979--983.

\bibitem{Sason06}
I.~Sason and S.~Shamai~(Shitz), ``{Performance analysis of linear codes under
  maximum-likelihood decoding: A tutorial},'' \emph{Foundations and Trends in
  Commun. and Inform. Theory}, vol.~3, pp. 1--222, June 2006.

\bibitem{Ma04}
X.~Ma and L.~Ping, ``Coded modulation using superimposed binary codes,''
  \emph{IEEE Trans. Inf. Theory}, vol.~50, pp. 3331--3343, Dec. 2004.

\bibitem{Liang14}
C.~Liang, K.~Huang, X.~Ma, and B.~Bai, ``{Block Markov superposition
  transmission with bit-interleaved coded modulation},'' \emph{IEEE Commun.
  Lett.}, vol.~18, pp. 397--400, Mar. 2014.

\bibitem{Dolecek09}
L.~Dolecek, P.~Lee, Z.~Zhang, V.~Anantharam, B.~Nikolic, and M.~Wainwright,
  ``{Predicting error floors of structured LDPC codes: Deterministic bounds and
  estimates},'' \emph{IEEE J. Sel. Area. Commun.}, vol.~27, pp. 908--917, Aug.
  2009.

\bibitem{Kavcic03}
A.~Kavcic, X.~Ma, and M.~Mitzenmacher, ``{Binary intersymbol interference
  channels: Gallager codes, density evolution, and code performance bounds},''
  \emph{IEEE Trans. Inf. Theory}, vol.~49, pp. 1636--1652, July 2003.

\end{thebibliography}
%%
%% where we here have assume the existence of the files
%% definitions.bib and bibliofile.bib.
%% BibTeX documentation can be obtained at:
%% http://www.ctan.org/tex-archive/biblio/bibtex/contrib/doc/
%%
%%
%%
%% Or manual references (pay attention to consistency!):

%\balance

\end{document}